\documentclass[twocolumn,prl,preprintnumbers,superscriptaddress,showpacs]{revtex4}

\usepackage{graphicx}
\usepackage{amsmath,amsfonts,amssymb,amsxtra}
\usepackage{units}
\usepackage{booktabs}
\usepackage{dcolumn}
\usepackage{sidecap}
\usepackage{rotating}

\makeatletter
\DeclareRobustCommand{\chemical}[1]{%
  {\(\m@th
   \edef\resetfontdimens{\noexpand\)%
       \fontdimen16\textfont2=\the\fontdimen16\textfont2
       \fontdimen17\textfont2=\the\fontdimen17\textfont2\relax}%
   \fontdimen16\textfont2=2.7pt \fontdimen17\textfont2=2.7pt
   \mathrm{#1}%
   \resetfontdimens}}
\DeclareRobustCommand{\bchemical}[1]{%
  {\(\m@th
   \edef\resetfontdimens{\noexpand\)%
       \fontdimen16\textfont2=\the\fontdimen16\textfont2
       \fontdimen17\textfont2=\the\fontdimen17\textfont2\relax}%
   \fontdimen16\textfont2=2.7pt \fontdimen17\textfont2=2.7pt
   \mathbf{#1}%
   \resetfontdimens}}
\makeatother

\newcommand{\tn }{$T_\text{N}$}

\newcommand{\smo}{SmFeO$_{3}$}

\newcommand{\idmi}{inverse Dzyaloshinskii-Moriya interaction}
\newcommand{\ts }{$T_\text{SR}$}

\begin{document}

\title{k=0 magnetic structure and absence of ferroelectricity in \smo}

\author{C.-Y. Kuo}
\author{Y. Drees }
\affiliation{Max-Planck-Institute for Chemical Physics of Solids, N\"{o}thnitzer Str. 40, 01187 Dresden, Germany}
\author{M. T. Fern\'{a}ndez-D\'{\i}az}
\affiliation{Institut Laue-Langevin, 38042 Grenoble, France}
\author{L. Zhao}
\affiliation{Institute of Physics. Academia Sinica, Taipei 11529, Taiwan}
\author{L. Vasylechko}
\affiliation{Max-Planck-Institute for Chemical Physics of Solids, N\"{o}thnitzer Str. 40, 01187 Dresden, Germany}
\affiliation{Lviv Polytechnic National University, 12 Bandera St., 79013 Lviv, Ukraine}
\author{D. Sheptyakov}
\affiliation{Laboratory for Neutron Scattering and Imaging, Paul Scherrer Institut, CH-5232 Villigen PSI, Switzerland}
\author{A. M. T. Bell}
\affiliation{HASYLAB at DESY, Notkestrasse 85, 22607 Hamburg, Germany}

\author{T. W. Pi}
\author{H.-J. Lin}
\affiliation{National Synchrotron Radiation Research Center (NSRRC), 101 Hsin-Ann Road, Hsinchu 30077, Taiwan}
\author{M.-K. Wu}
\affiliation{Institute of Physics. Academia Sinica, Taipei 11529, Taiwan}
\author{E. Pellegrin}
\author{S. M. Valvidares}
\affiliation{CELLS-ALBA Synchrotron Radiation Facility, Carretera BP 1413, km 3.3, E-08290
Cerdanyola del Vall`es, Barcelona, Spain}
\author{Z. W. Li}
\author{P. Adler}
\author{A. Todorova}
\author{R. K\"{u}chler}
\author{A. Steppke}
\author{L. H. Tjeng}
\author{Z. Hu }
\author{A. C. Komarek}
\email{Alexander.Komarek@cpfs.mpg.de}%
\affiliation{Max-Planck-Institute for Chemical Physics of Solids, N\"{o}thnitzer Str. 40, 01187 Dresden, Germany}
\date{\today}

\begin{abstract}
\smo\ has attracted considerable attention very recently due to the reported multiferroic properties above room-temperature.
We have performed powder and single crystal neutron diffraction as well as complementary polarization dependent soft X-ray absorption spectroscopy measurements on floating-zone grown \smo\ single crystals in order to determine its magnetic structure. We found a k=0 G-type collinear antiferromagnetic structure that is not compatible with \idmi\ driven ferroelectricity.
While the structural data reveals a clear sign for magneto-elastic coupling at the N\'{e}el-temperature of $\sim$675~K, the dielectric measurements remain silent as far as ferroelectricity is concerned.
\end{abstract}

\maketitle

The discovery of magnetism induced ferroelectricity has renewed the interest on multiferroic materials due to the enhanced magnetoelectric interaction in these materials that makes them very interesting for technical applications \cite{allg,allgB}.
Recently, it has been found, that ferroelectricity can arise from some special types of magnetic structures inducing sizeable magnetoelectric effects and the ability to switch the electric polarization by an applied magnetic field (and vice versa). In these recently studied materials magnetic frustration induces a complex magnetic structure with cycloids or spirals \cite{allg}.
However, also the ordering temperatures need to be above room-temperature in order to make these materials interesting for technical applications.
Very recently, the discovery of ferroelectric polarization in \smo\ has been reported below \tn$\sim$670~K.
The origin of this ferroelectric polarization is highly debated \cite{smoA,smoB,smoC}.
A further spin-reorientation transition occurs at \ts$\sim$480~K in this material
without having any noticeable effect on the ferroelectric polarization.
In the initial publication an \idmi\ based mechanism has been reported to be the driving force of the ferroelectric properties of \smo\ \cite{smoA}.
The underlying k=0 magnetic structure was not directly measured but calculated by \emph{ab initio} calculations \cite{smoA}.
However, it has been demonstrated that this calculated k=0 magnetic structure with magnetic ions located at inversion
centers can not be responsible for a spin-orbit-coupling driven ferroelectric polarization by \textbf{S}$_i$~$\times$~\textbf{S}$_j$ in this material
since inversion symmetry will not be broken \cite{smoB}.
Within this context, an alternative mechanism based on $J$\textbf{S}$_i$~$\bullet$~\textbf{S}$_j$ exchange-striction has been proposed to be responsible for the ferroelectric polarization in \smo\ \cite{smoC}.

\begin{figure}[!t]
\begin{center}
\includegraphics*[width=1\columnwidth]{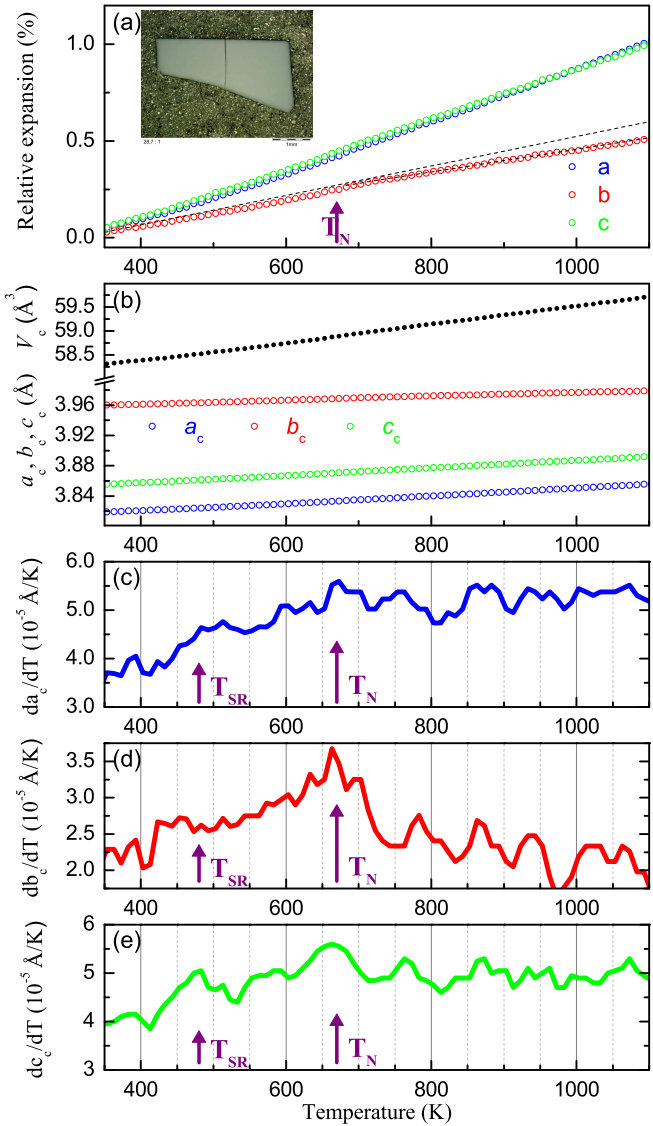}
\end{center}
\caption{(color online) Results of our synchrotron radiation powder X-ray diffraction measurements of \smo\ obtained at beamline B2 of DORIS-III at DESY ($\lambda$~=~0.538163~\AA, space group \emph{Pbnm}). (a) Relative expansion $r(a)=(a(\text{T})-a(\text{300 K}))/a(\text{300 K})$. In the inset a polarization microscope image of a single crystalline sample is shown. (b) Unit cell volume and pseudocubic lattice parameters $V_c=V/4$, $a_{c}=a/\sqrt{2}$, $b_{c}=b/\sqrt{2}$, $c_{c}=c/2$. (c-e) The derivative d$a_{c}$/dT of the pseudocubic lattice parameters (averaged over 4 data points). All dashed lines are guide to the eyes.}
\label{fig1}
\end{figure}
Here, we report the experimentally observed magnetic structure of \smo\ and re-analyze the ferroelectric properties of \smo.
A large reddish \smo\ single crystal with  \tn$\sim$675~K has been grown at a \emph{Crystal systems Corp.} 4-mirror optical floating zone furnace. The high sample quality of this insulating reddish orange single crystal has been confirmed by EDX, Laue and X-ray diffraction techniques as well as by magnetization and M\"{o}ssbauer spectroscopy measurements, see the Supplementary Materials \cite{suppl}. In the inset of Fig.~\ref{fig1}~(a) a polarization microscope image is shown that indicates - together with our Laue diffraction analysis - that our \smo\ crystals are single domain single crystals. No impurity phases are visible in highly accurate synchrotron radiation powder X-ray diffraction measurements that have been performed at beamline B2 of DORIS-III at DESY.
The lattice parameters, unit cell volume and relative expansion of the lattice parameters of \smo\ (\emph{Pbnm} setting with $a$~$<$~$b$~$<$~$c$) are shown in Figs.~\ref{fig1}~(a,b). The $b$-lattice parameter exhibits a small anomalous kink around \tn\ that is indicative for magneto-elastic coupling at the magnetic ordering temperature \tn. The other lattice parameters exhibit much less pronounced anomalies at \tn\ that are even only barely visible in the derivatives of the lattice parameters, see Figs.~\ref{fig1}~(c-d).

\par
\begin{figure}[!tl]
\begin{center}
\includegraphics*[width=1\columnwidth]{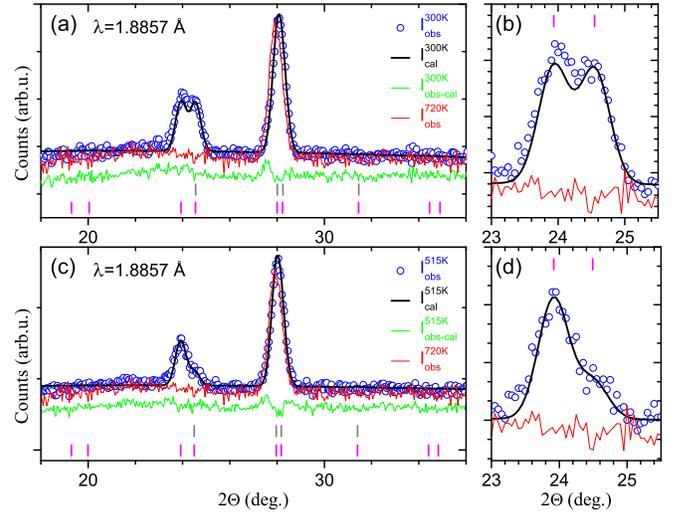}
\end{center}
\caption{(color online) Powder neutron diffraction measurements of \smo\ at (a-b) 300~K and at (c-d) 515~K. \emph{Blue circles}: measured intensities, \emph{black line}: Rietveld fit, \emph{green line}: $I_{obs}-I_{cal}$, \emph{red line}: measured intensities in the paramagnetic phase at 720~K, \emph{grey/magenta bars}: nuclear/magnetic peak positions.  }
\label{fig3}
\end{figure}

 The highly neutron absorbing properties of the element Sm ($\sim$5900 barn for 2200 m/s neutrons) hampered the experimental determination of the magnetic structure of \smo. Here, we present two complementary neutron measurements where we were able to overcome these obstacles and measure the magnetic structure of \smo\ directly.
First, we have performed powder neutron diffraction measurements at comparably low neutron energies ($\lambda$=1.8857\AA) with a special sample geometry at the HRPT diffractometer (SINQ). We were able to overcome the highly neutron absorbing properties of Sm by filling the outer volume of a hollow vanadium cylinder with a special mixture of fine \smo\ powder that we "diluted" with fine aluminum powder in order to suppress the Sm absorption effects.
As can be seen in Fig.~\ref{fig3} we obtained qualitatively good powder neutron diffraction patterns at 300~K, 515~K and 720~K that could be easily refined with two additional phases of Al and V which do not interfere at all with the \smo\ magnetic signal and even barely with structural contributions. Therefore, a reliable Rietveld refinement of the magnetic structure of \smo\ could be performed.

\begin{figure}[!tr]
\begin{center}
\includegraphics*[width=1\columnwidth]{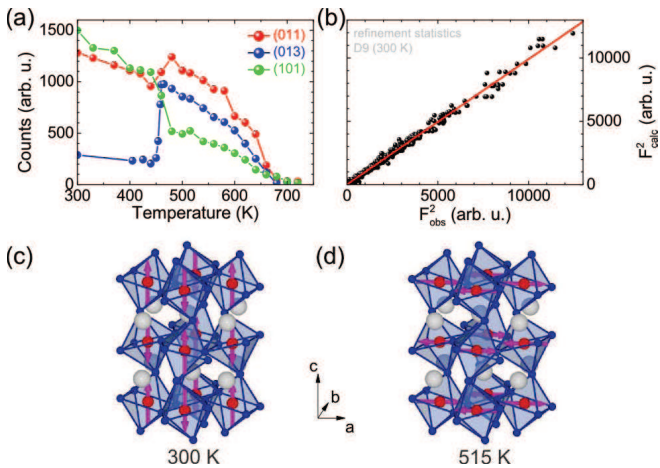}
\end{center}
\caption{(color online) (a) Temperature dependence of magnetic intensities measured on our \smo\ single crystal at the D9 diffrectometer. (b) Refinement statistics of our magnetic and crystal structure refinement of \smo\ at 300~K. (c,d) Magnetic structures of \smo\ below and above \ts\ respectively:  }
\label{fig3b}
\end{figure}

\begin{figure}[!t]
\begin{center}
\includegraphics*[width=1\columnwidth]{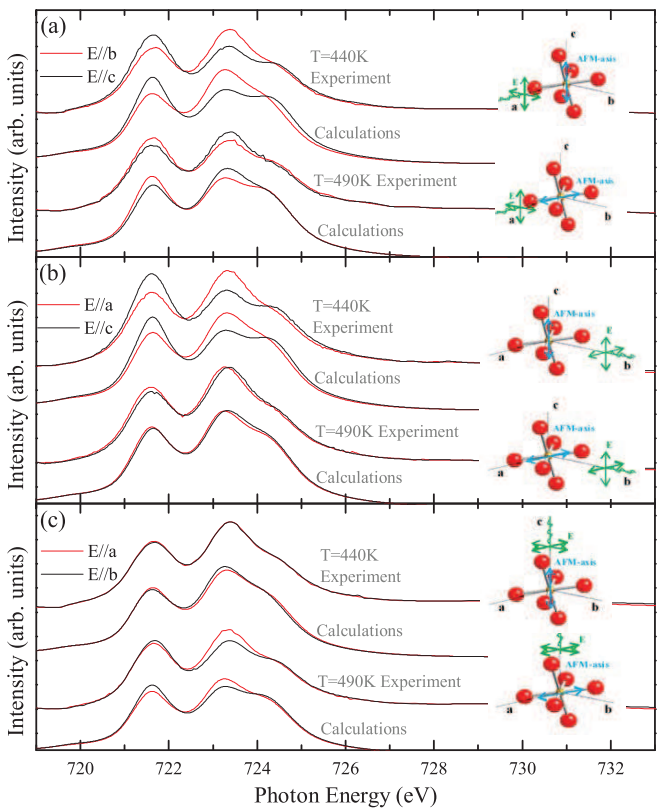}
\end{center}
\caption{(color online) Fe L$_{2,3}$~XAS spectra of \smo. (a-c) The linear polarization dependent XAS spectra measured above and below \ts\ with incident beam parallel to $a$-, $b$- and $c$-axis.}
\label{fig4}
\end{figure}

Since we measured only at temperatures above room-temperature, we neglected any Sm ordering  \cite{SmOrdering}.
As pointed out in great detail in Ref.~\cite{smoB} there are four different irreducible representations $\Gamma_1^+$, $\Gamma_2^+$, $\Gamma_3^+$ and $\Gamma_4^+$ corresponding to the following four spin configurations for the Fe ordering: A$_x$G$_y$C$_z$, G$_x$A$_y$F$_z$, F$_x$C$_y$G$_z$ and C$_x$F$_y$A$_z$ which correspond to the magnetic space groups \emph{Pbnm}, \emph{Pb'n'm}, \emph{Pbn'm'} and \emph{Pb'nm'} respectively \cite{Bertaut}.
Our neutron measurements at 300~K, 515~K and 720~K clearly show that there appears F$_x$C$_y$G$_z$-type and G$_x$A$_y$F$_z$-type antiferromagnetic ordering in \smo\ at 300~K and 515~K respectively. The Rietveld fits of the magnetic intensities are shown in Fig.~\ref{fig3} and the corresponding magnetic moments are listed in Table~I within the Supplementary Materials \cite{suppl}. Other spin configurations or incommensurate magnetic structures can be excluded for \smo.
We have also performed complementary single crystal neutron diffraction measurements. By choosing an optimized small sample geometry and high incident neutron energies we were able to perform single crystal neutron diffraction measurements at the D9 diffractometer (ILL).
The temperature dependence of some prominent magnetic intensities are shown in Fig~\ref{fig3b}~(a) visualizing the spin-reorientation transition.
The atomic positions derived from this measurement are very close to the atomic positions derived from a complementary single crystal X-ray diffraction measurement that has been performed on a \emph{Bruker D8 VENTURE} X-ray diffractometer as well as with values given in literature \cite{elec}, thus, proving the high reliability of our neutron measurements, see Table~I within the Supplementary Materials \cite{suppl}. The comparably good refinement statistics and R-values of the structure refinement of this single crystal neutron measurement are shown in Fig.~\ref{fig3b}~(b) and Table~I within the Supplementary Materials \cite{suppl}.
Finally, we were able to determine the magnetic structure of \smo\ and observe a collinear k=0 antiferromagnetic structure, see Fig.~\ref{fig3b}~(c,d). The detection of very tiny canted magnetic moments is beyond the scope of these measurements. As pointed out in Ref.~\cite{smoB} for \smo\ (with magnetic ions located at inversion centers), k=0 magnetic structures are not compatible with an \idmi\ induced electric polarization.

\par
The antiferromagnetic properties of \smo\ were also studied by linear polarization dependent Fe-L edge X-ray absorption spectra (XAS), conducted at 08B beam line of the National Synchrotron Radiation Research Center (NSRRC) in Taiwan. The spectra were recorded with the total electron yield mode using Fe$_2$O$_3$ for calibration. X-ray magnetic linear dichroism (XMLD) is the difference in cross section for light polarized perpendicular or parallel to the magnetic moment and is well known to be sensitive to the spin direction of AFM systems \cite{xmcdA,xmcdB,xmcdC}. We have measured the polarization dependent Fe-L$_2$ XAS spectra at 440~K and 490~K with the Poynting vector of the light being parallel to the $a$-, $b$- and $c$-axis shown in Fig.~\ref{fig4}~(a-c). We observe a considerable size of the XMLD signals between the electric field $E||b$ and $E||c$ in Fig.~\ref{fig4}~(a), between $E||a$ and $E||c$ in Fig.~\ref{fig4}~(b), but nearly no difference between $E||a$ and $E||b$ in Fig.~\ref{fig4}~(c). The sign of the XMLD signals is reversed when going from 440~K to 490~K, see Fig.~\ref{fig4}(a-b). This is similar to the previous study of the Morin transition of Hematite \cite{xmcdA} revealing a rotation of the spin orientation across \ts. To extract the orientations of the AFM axes we have simulated the experimental spectra using configuration interaction cluster calculations \cite{tanaka}. The calculated spectra are shown in Fig.~\ref{fig4}~(a-c) and the parameters used in our calculation are listed in Ref.~\cite{CY}. The corresponding FeO$_6$ cluster considered in our calculations is also shown in the right part of each figure. One can see that the experimental spectra are nicely reproduced by the calculated spectra with spins parallel to $c$- and $a$-axis at 440~K and 490~K respectively, thus, corroborating the collinear magnetic structure obtained in our neutron measurements.

\par Also our Fe-L$_{2,3}$ X-ray magnetic circular dichroism (XMCD) spectra as well as our M\"{o}ssbauer spectroscopy measurements are fully consistent with the fact that there is only one Fe$^{3+}$ species in \smo, see Supplementary Materials \cite{suppl}.

\begin{figure}[!t]
\begin{center}
\includegraphics*[width=1\columnwidth]{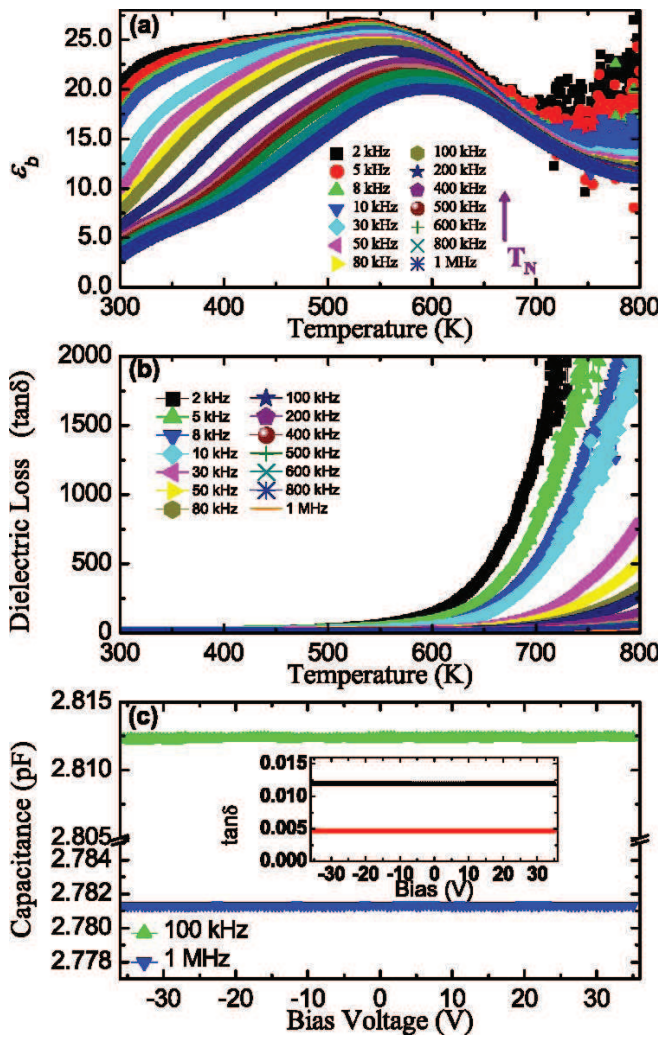}
\end{center}
\caption{(color online)  Dielectric characterization of \smo\ with the electric field along the $b$-axis at different testing frequencies. (a) The temperature-dependent dielectric constant $\varepsilon_b$. (b) The tangent loss tan$\delta$. (c) The $C$-$V$ curve of \smo\ at room temperature with the corresponding loss data in the inset.}
\label{fig5}
\end{figure}	
\par Finally, we measured the anisotropic dielectric properties of single crystalline thin plates of \smo. The capacitance was measured over a range of frequencies with an excitation level of 1~V, while the temperature was swept at a slow warming or cooling rate (1-2~K/min).
As shown in Fig.~\ref{fig5}~(a,b) the temperature-dependent dielectric constant with electric field along $b$-axis, $\varepsilon_b(T)$, shows only a broad hump with strong frequency-dependence below $\sim$600~K. The dielectric loss, tan$\delta$, rises strongly as the temperature increases. All samples are insulators at room temperature and become slightly conductive at high temperatures (several k$\Omega$ at 800~K).
No apparent anomalies could be observed in $\varepsilon_b(T)$ around \tn. If an intrinsic ferroelectric transition occurs at \tn, the corresponding anomalies
should be observable in both $\varepsilon_b$ and tan$\delta$, irrespective of testing frequencies. Also $\varepsilon_a(T)$ and $\varepsilon_c(T)$ exhibit no anomalies at \tn.
Complementary capacitance-voltage ($C$-$V$) measurements were carried out for all our samples at room temperature. Fig.~\ref{fig5}~(c) shows a typical $C$-$V$ curve with the electric field applied along the $b$-axis. No hysteresis could be observed for \smo\ within the experimental resolution ($<10^{-4}$) .

\par Hence, we can exclude the existence of ferroelectricity in \smo.
We interpret the observations in Ref.~\cite{smoA} differently and suggest that strain could be induced by magnetoeleastic coupling at \tn\ which then would be responsible for an artificial observation of a pyrocurrent in $b$-direction at \tn. Indeed, our synchrotron radiation powder X-ray diffraction measurements reveal anomalies predominantly of the $b$-lattice parameter of \smo\ at \tn, see Fig.~1~(d).
The hysteresis loop reported to occur at 300~K in Ref.~\cite{smoA} may then be attributed to leakage currents \cite{scott} which is absent in our experiment. Perhaps this is related to the different lossy character of flux-grown \cite{smoA} and floating zone grown single crystals.

The absence of ferroelectric properties in \smo\ is also consistent with the k=0 magnetic structure that we observed. This has to be contrasted with the case of other isostructural multiferroic materials, like TbMnO$_3$, where non-collinear chiral magnetic structures have been observed \cite{tb}.
We note that for a G-type antiferromagnetic rare-earth orthoferrite $R$FeO$_3$, the electric polarization induced by exchange striction is known to occur only below the rare earth magnetic ordering temperature which is two orders of magnitude lower than \tn\ \cite{gdfeo}.  If exchange striction would be an important mechanism in \smo\ one would expect to see also a pyrocurrent signal when the magnetic structure exhibits distinct changes at \ts\ which is not experimentally observed \cite{smoA}.
Finally, we would like to remark that magneto-elastic effects are not only present in prototypical multiferroic materials like BiFeO$_3$ \cite{bifeoA,bifeoB} but (across the doping series Bi$_{1-x}$La$_{x}$FeO$_3$ \cite{bifeoC}) also in non-ferroelectric centrosymmetric materials like LaFeO$_3$ \cite{lafeoA,lafeoB}.
Our findings suggest that magneto-elastic effects may also lead to an artificial observation of pyrocurrents and, hence, magnetoelastic coupling can easily be misinterpreted as a ferroelectric response.

\subsection{Acknowledgements} We thank D.~I.~Khomskii, M.~W.~Haverkort and A.~Tsirlin for helpful discussions.
We thank H.~Borrmann and his team for X-ray diffraction measurements. We thank U.~Burkhardt and his team for EDX measurements.
This work is partially based on experiments performed at the Swiss spallation neutron source SINQ, Paul Scherrer Institute, Villigen, Switzerland. XMCD experiments were performed at the BL29 Boreas beamline at the ALBA Synchrotron Light Facility with the collaboration of ALBA staff.

\newpage

\begin{table*}
\centering
\begin{tabular}{c}

{\large \textbf{Supplemental Materials for}}\\
{\large \textbf{k=0 magnetic structure and absence of ferroelectricity in \smo}}\\

\end{tabular}
\end{table*}

\clearpage\newpage

\begin{figure}[!b]
\begin{center}
\includegraphics*[width=0.90\columnwidth]{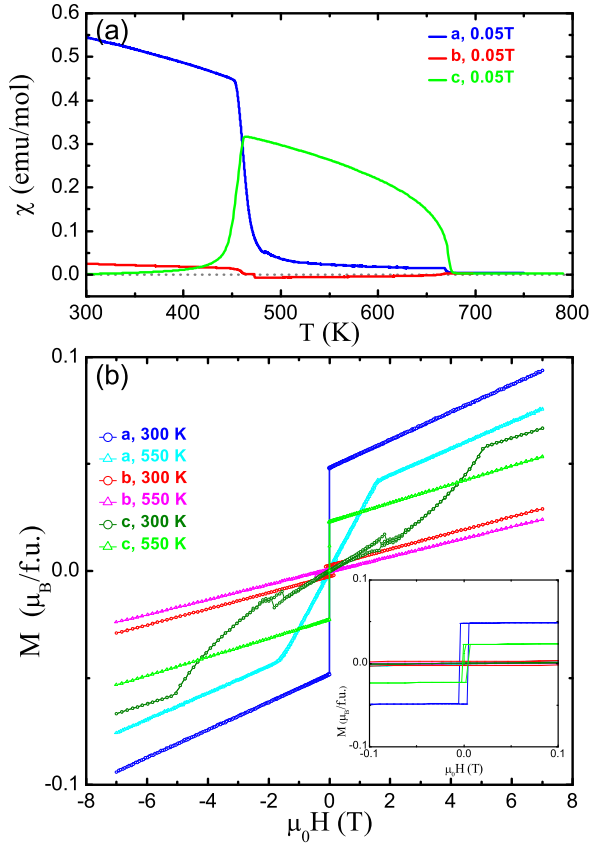}
\end{center}
\caption{(a) Magnetic susceptibility of \smo\ as a function of temperature for three different crystallographic directions ($\mu_{0}\cdot H$~=~0.05~T). (b) Magnetization measurements of \smo\ below \tn\ and below \ts.}
\label{figMag}
\end{figure}
\par The results of our single crystal and powder neutron diffraction as well of our single crystal X-ray diffraction measurements are listed in Table~\ref{tab1}. The crystal structure of \smo\ which been determined by single crystal X-ray diffraction using Mo-K$_{\alpha}$ radiation is close to literature values \cite{elec}. The magnetic structure of \smo\ could be obtained from powder neutron diffraction measurements that have been performed at the HRPT diffractometer at SINQ ($\lambda$~=~1.8857~\AA). Complementary single crystal neutron diffraction measurements have been performed at the D9 diffractometer at the ILL ($\lambda$~=~0.511~\AA). Crystal and magnetic structure of \smo\ could be obtained at the same time in this single crystal neutron measurement. The structural parameters (including ADP) are extremely close to the ones obtained by single crystal X-ray diffraction proving the high reliability of our room-temperature single crystal neutron data and that we were able to cope with the highly neutron absorbing properties of the Sm-ions. The high temperature single crystal neutron data was measured using a furnace and much less reflections have been collected leading to less accurate results. The detection of extremely tiny ferrolectric distortions of the Fe-ions or any small cantings of magnetic moments is beyond the scope of all these measurements.

\begin{table}[!h]
\centering{ {\footnotesize
\begin{ruledtabular}
\begin{tabular}[t]{c|ccc }
  & \multicolumn{3}{c}{single crystal diffraction} \\
method           &    X-ray     &        neutron   & neutron           \\
Temp.            &    $\sim$300~K     &      300~K    & 500~K        \\
reflections      &   64479        &        983    &   246   \\
redundancy      &    26.3        &         2.28   & 1.58       \\
R$_{int}$(\%)      &   2.20         &        2.85   & 5.53       \\
GoF               &   1.39        &          - & - \\
R/R$_w$(\%)      &  1.70/4.73      &       6.23/6.65 &  27.3/6.64             \\
\hline
\hline
x(Sm1)        &{ 0.986950(2)}  &{ 0.98662(2) }  &   { 0.99547(8) }    \\
y(Sm1)        &{ 0.056695(2)}  &{  0.05675(1)  }   &   { 0.05657(0) }     \\
U(Sm1)        &{ 0.005118(2)}  &{ 0.00604(4) }   &   { 0.017(4) }     \\
U(Fe1)       &{ 0.004207(5)}  &{ 0.00348(1) }    &   { 0.017(4) }    \\
x(O1)            &{ 0.09424(3)}  &{ 0.09476(2) }   &   { 0.04999(11) }     \\
y(O1)            &{ 0.47096(3)}  &{ 0.47109(1) }    &   { 0.48585(18) }    \\
U(O1)            &{ 0.00619(2)}  &{ 0.00626(1) }    &   { 0.017(4) }    \\
x(O2)            &{ 0.69981(2)}  &{ 0.69947(1) }    &   {  0.72279(7) }     \\
y(O2)            &{ 0.29896(2)}  &{ 0.29913(1) }    &   { 0.28984(4) }    \\
z(O2)            &{ 0.04933(2)}  &{ 0.04971(0)  }    &   { 0.02830(10) }     \\
U(O2)            &{ 0.00642(2)}  &{ 0.00609(1) }    &   { 0.017(4) }    \\
\hline
\hline
AFM config.          &   &  F$_x$C$_y$G$_z$ &  G$_x$A$_y$F$_z$  \\
M$_x$ ($\mu_B$)                &        &   -       &  2.836(112)  \\
M$_y$ ($\mu_B$)               &            &   -    &   -    \\
M$_z$ ($\mu_B$)                &        &  3.692(53)     &   -       \\
\hline\hline
  & \multicolumn{3}{c}{powder diffraction} \\
method              & & neutron   & neutron  \\
Temp.               & & 300~K   & 515~K \\
R$_{mag.}$(\%)      &        &   23.3         &     22.0            \\
\hline \hline
AFM config.          &  &      F$_x$C$_y$G$_z$ &  G$_x$A$_y$F$_z$     \\
M$_x$ ($\mu_B$)                &  &      -               &  3.020(110)        \\
M$_y$ ($\mu_B$)               &  &      -               &  -                  \\
M$_z$ ($\mu_B$)                &  &      3.759(71)       &  -                      \\
\hline\hline
 &  \multicolumn{3}{c}{lattice parameters} \\
Temp. &	$a$ (\AA) & $b$ (\AA) & $c$ (\AA) \\
301~K &	5.39827(9) & 5.59872(9) & 7.7071(1) \\
493~K & 5.4089(1)  & 5.6054(1)  & 7.7236(2) \\
503~K & 5.4095(1)  & 5.6057(1)  & 7.7251(2) \\
513~K & 5.4102(1)  & 5.6063(1)  & 7.7256(2) \\
523~K & 5.4109(1)  & 5.6064(1)  & 7.7264(2) \\
\end{tabular}
\end{ruledtabular}
}} \caption{\label{tab1} Crystal and magnetic structure of \smo\ derived by single crystal X-ray, powder neutron and single crystal neutron diffraction measurements (space group $Pbnm$,
z(Sm1)=0.25, x(Fe1)=0, y(Fe1)=0.5, z(Fe1)=0, z(O1)=0.25).   Finally, the corresponding lattice parameters obtained by synchrotron radiation powder X-ray diffraction at beamline B2 are listed.}
\end{table}

\par We measured the magnetic susceptibility of our \smo\ single crystals using a SQUID-VSM magnetometer. As can be seen in Fig.~\ref{figMag}, the findings in Ref.~\cite{smoA} can be reproduced by our \smo\ single crystals. However, our single crystals have roughly 5~K higher magnetic ordering temperature \tn$\sim$675~K than that in Ref.~\cite{smoA}.
The $M$-$H$ curve in $b$-direction that was not shown in Ref.~\cite{smoA} also exhibits a step-like jump at little higher coercive fields, i.e. at about $\sim$0.14~T instead of at about $\sim$0.0035~T like for the $a$-direction. This also proves that we did not simply measure the $a$-direction effect in a slightly misaligned sample.
\begin{table}[!t]
\centering{ {\footnotesize
\begin{ruledtabular}
\begin{tabular}[t]{rl||rl }
isomer shift: &   0.3672(6)~mm/s  & hyperfine field:  & 50.88(1)~T \\
q. splitting:  &  -0.104(1)~mm/s   & line width:  & 0.237(2)~mm/s \\
\end{tabular}
\end{ruledtabular}
}} \caption{\label{tab0} Hyperfine parameters obtained by least square fits of the M\"{o}ssbauer spectrum of our orange \smo\ samples (q. splitting: abbreviation for quadrupole splitting). Note that in contrast to Ref.~\cite{smoMoess} the isomer shift is given relative to alpha-iron. The quadrupole splitting value here is twice the coupling constant given in Ref.~\cite{smoMoess}}
\end{table}

\par
\begin{figure}[!t]
\begin{center}
\includegraphics*[width=1\columnwidth]{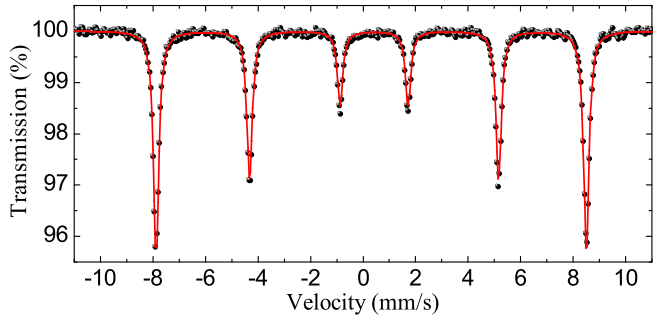}
\end{center}
\caption{M\"{o}ssbauer spectrum of our \smo\ sample at room-temperature; \emph{black dots}: measured data, \emph{red line}: calculated transmission.}
\label{figS1}
\end{figure}

 Complementary M\"{o}ssbauer spectroscopy measurements indicate with high accuracy solely one Fe$^{3+}$ species within our orange \smo\ powder samples that were obtained by crushing parts of our single crystals. The spectrum, which was obtained at room temperature with a standard spectrometer operating in the constant acceleration mode, can be well fitted with only one sextet consistent with the fact that there is only one crystallographic Fe site, see Fig.~\ref{figS1}. The deduced hyperfine parameters are listed in Table~\ref{tab0} and are consistent with published data in literature \cite{smoMoess}. The very narrow line width of $\sim$0.237~mm/s corroborates the high quality of our \smo\ crystals and is in agreement with the collinear antiferromagnetic spin structure derived from the neutron diffraction study. In this case a well defined magnetic hyperfine field and orientation between hyperfine field and electric field gradient results in a sharp pattern. On the other hand, incommensurate spin structures could lead to line broadening due to variations in the orientation between the electric field gradient and the hyperfine field and/or in the  size of the hyperfine field. The spin reorientation transition at \ts\ in \smo\ is reflected as a change of the quadrupole coupling \cite{smoMoess}.

\par
\begin{figure}[!t]
\begin{center}
\includegraphics*[width=1\columnwidth]{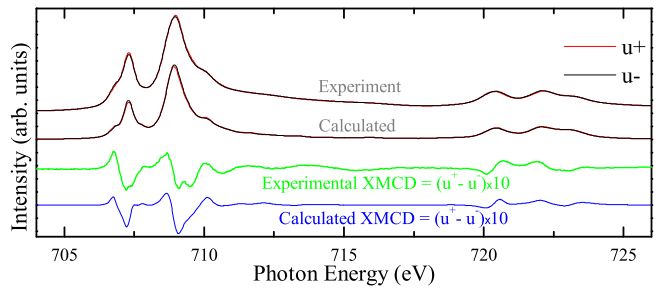}
\end{center}
\caption{Experimental and calculated circular polarization dependent XAS and XMCD spectra at 380~K with incident beam parallel to the $a$-axis.}
\label{figXMCD}
\end{figure}

We have also measured the Fe-L$_{2,3}$ X-ray magnetic circular dichroism (XMCD) spectra at the BL29 Boreas beamline at the CELLS-ALBA synchrotron radiation facility (Barcelona, Spain). The XMCD spectrum shown in Fig.~\ref{figXMCD} is the difference between the signals with photon spin parallel (u+) and antiparallel (u-) to the magnetic field. Its shape is similar to the previous study \cite{smoA} but the size of the our XMCD is enhanced by more than factor two due to high magnetic field in this work. We can well reproduce the XMCD shape at both, the L$_3$ and L$_2$ edges. Although there are four Fe atoms in the unit cell of \smo, our calculations show that each of these four sites gives the same XMCD spectrum. Therefore, the four major dichroism peaks do not arise from four sets of the spin subsystem as concluded in Ref.~\cite{smoA}, but from the local symmetry of each single Fe ion.

\end{document}